\documentclass{aastex62}
\shorttitle{Energy Partition of Magnetic Reconnection}
\shortauthors{Hoshino}

\begin{document}

\title{Energy Partition between Ion and Electron of Collisionless Magnetic Reconnection}

\correspondingauthor{Masahiro Hoshino}
\email{hoshino@eps.s.u-tokyo.ac.jp}

\author[0000-0002-1818-9927]{Masahiro Hoshino}
\affil{Department of Earth and Planetary Science,
The University of Tokyo, Tokyo 113-0033, Japan}

\begin{abstract}
The plasma heating during collisionless magnetic reconnection is investigated
using particle-in-cell simulations. We analyze the time evolution of the plasma
temperature associated with the motion of the reconnecting flux tube, 
where the plasma temperature is defined as second-order moment of velocity distribution
function in the simulation frame/in the center of the flux tube frame, and we show
that the plasma heating during magnetic reconnection can be separated into
two distinct stages: the nonadiabatic heating stage, in which the magnetic field
lines are just reconnecting in the X-type diffusion region, and the adiabatic
heating stage, in which the flux tube is shrinking after two flux tubes merge.
During the adiabatic heating stage, the plasma temperature
$T$ can be approximated by $TV^{\gamma-1}=const.$, where $\gamma=5/3$ is
the specific heat, and $V$ is the volume of the flux tube.
In the nonadiabatic heating stage, we found numerically that the ratio of the
increment of the ion temperature to that of the electron temperature can be
approximated by $\Delta T_i/\Delta T_e \approx (m_i/m_e )^{1/4}$, where $m_i$ and
$m_e$ are the ion and electron mass, respectively.  We also present a theoretical
model based on a magnetic-diffusion-dominated
reconnection to explain the simulation result.
\end{abstract}

\section{Introduction}
High energy astrophysical phenomena such as planetary magnetospheres,
solar flares, pulsar wind nebulae, astrophysical jets and magnetars often
contain the high magnetic energy density that exceeds the plasma energy
density \citep[e.g.,][]{Lyubarsky01,Spruit10,Uzdensky11,Cerutti13,Sironi14}.
By releasing the magnetic field energy into energetic particles and
electromagnetic radiations, dramatic magnetoactive phenomena
can be observed \citep[e.g.,][]{Birn12,Hoshino12,Blandford17}.
While magnetic reconnection is believed to be the major energy conversion
process in these sources, energy partition between hadronic and leptonic
components during reconnection remains a fundamental quest in hot and
dilute collisionless plasma. Non-equipartition of energy between
ion and electron is often a key issue to discuss the dynamical evolution
of the astrophysical phenomena.
We discuss the energy partition during collisionless magnetic reconnection.  

It is observationally known that ions are preferentially heated compared to
electrons during magnetic reconnection in the earth's magnetotail, and that
the ion temperature is several times greater than the electron one
\citep{Baumjohann89,Wang12,Eastwood13}.  
Conversion of energy into plasma heat has also been observationally
investigated in other types of reconnection in the solar wind and at
the earth's magnetopause, and it is found that the energy transfer to ions
is larger than that to electrons \citep{Phan13}. In addition to space
observations, kinetic simulations using the particle-in-cell
(PIC) method also suggest that the energy gain of ions
is larger than that of electrons, and that energy transfer to ions can
increase as the mass ratio of ions to electrons increases
\citep[e.g.,][]{Shay14,Haggerty15}.  The mechanism of energy conversion
from stored magnetic energy into ion and electron heating is a principal
topic of reconnection, but the energy partition remains under debate.

One of the difficulties in understanding the energy partition during reconnection
is that the ion and electron temperatures are not uniform in space and time.
Plasmas are quickly heated when the upstream plasma just enters the
magnetic diffusion region located at the X-type point, and some of the
accelerated high-energy particles are ejected outward along the magnetic
field lines \citep[e.g.][]{DeCoster79,Birn12,Egedal13}; 
their free energies, stored in the form of beams,
are then released in the separatrix away from the X-type point
\citep[e.g.][]{Hesse01,Hoshino01,Birn12}. 
In addition, the magnetic field lines reconnected at the diffusion
region are transported downstream into the reconnection exhaust,
\citep[e.g.,][]{Hoshino98,Lottermoser98,Drake09} and the plasma temperatures
increase further owing to plasma compression \citep[e.g.,][]{Imada05}.

To date, the plasma heating process and its energy partition mechanism have been
discussed by comparing the temperatures upstream of the reconnection
region with those of the downstream/exhaust region \citep{Eastwood13,Phan13,
Shay14,Haggerty15,Shay18}.  In this letter, however, we analyze the plasma
temperature associated with the motion of the reconnecting magnetic flux tube
in order to distinguish the adiabatic and nonadiabatic heating processes.
If no energy dissipation due to the nonadiabatic process exists, the plasma
entropy $p/\rho^{\gamma}$ along the plasma flow, i.e., along the so-called
Lagrangian plasma flow, should be conserved.  

It is usually not easy to analyze the evolution of the gas pressure and
density in Lagrangian flow, but under the frozen-in condition between
magnetic field lines and plasmas, Lagrangian flow can coincide with the motion
of the magnetic field lines outside of the diffusion region, 
and the motion of the magnetic field can be easily
described using the contour of the vector potential in a two-dimensional
system in space.  Using this simple idea, we define the
flux tube volume and plasma temperature integrated over the magnetic flux tube,
and we show that the time evolution of the ion and electron temperatures
can be separated into two processes:  effective Ohmic 
heating in the diffusion region and an adiabatic compression process associated with
the outward plasma convection in the reconnection exhaust.  

\section{Thermodynamical Properties of Magnetic Flux Tube}
\begin{figure}
\plotone{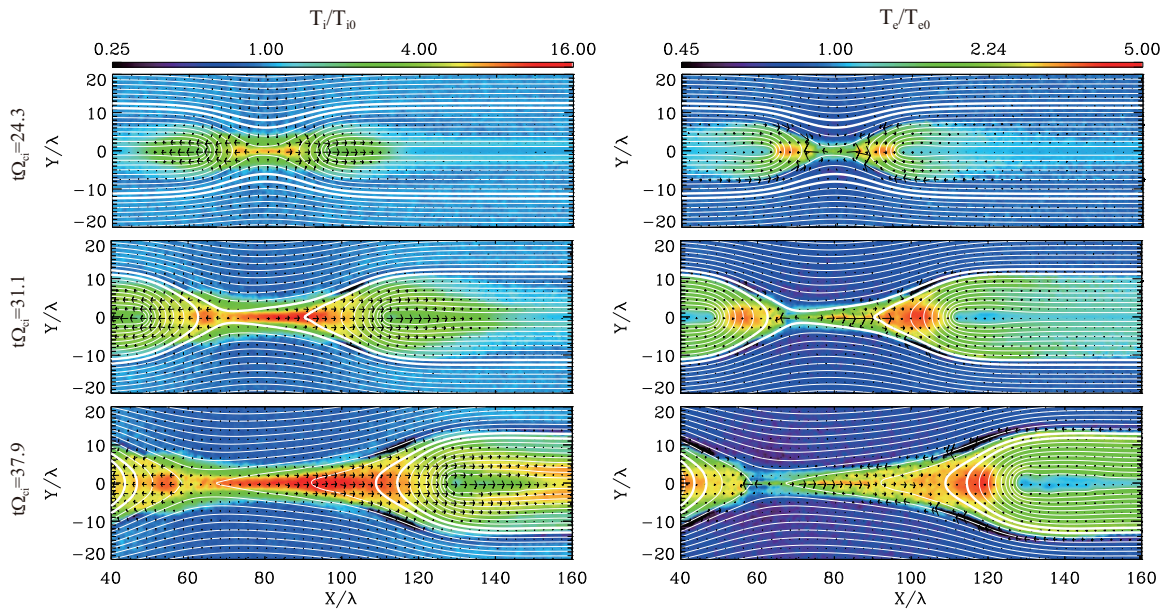}
\caption{Time evolution of collisionless magnetic reconnection. Left and right panels show the ion and electron temperatures, respectively.  The white lines represent magnetic field lines, and the thick white lines show 
a typical magnetic flux tube moving with the plasma flow.
Black arrows indicate the flow vectors in arbitrary units.  Top to bottom:
the stage before two flux tubes are reconnected, the reconnection stage, and the stage after two flux tubes merge.}
\end{figure}

Figure 1 shows the time evolution of magnetic reconnection obtained by
a two-dimensional PIC simulation with a periodic boundary in the $x$ direction and
conducting walls for the upper and lower boundaries at $y= \pm 40 \lambda$,
where $\lambda$ is the thickness of the initial plasma sheet.  The total system
size is $160 \lambda \times 80 \lambda$, and the grid size is $6400 \times 3200$.
Part of the simulation domain is depicted.  The spatial scale is normalized
by the thickness of the plasma sheet $\lambda$.  The mass ratio of ions to electrons
is $m_i/m_e=400$, and Harris equilibrium with the same ion and electron temperature
$T_{i0}=T_{e0}$ is assumed.  The total number of particles is $2.2 \times 10^{10}$,
 i.e., $538 \times 2$ particles/grid cell in order to calculate
the high $\beta$ plasma sheet as accurately as possible. 
The ion and electron inertial lengths are $V_A/\Omega_{ci}=56.5$ and
$c/\Omega_{pe}=2.8$ grid points, respectively, and $\lambda$ has $40$ grid points.
The ion drift velocity is set to be the same as the ion thermal velocity.
The background plasma density in the upstream region is $5 \%$ of the maximum density
of the plasma sheet.

The color contour in Figure 1 shows the ion and electron temperatures normalized
by the initial temperature,
and the scales are indicated by the bars at the top of the panels.
The white lines are the magnetic field lines calculated from the contour of the
vector potential $A_z(x,y,t)$, and the thick white lines show the time evolution
of the flux tube.  We observe that the flux tube is transported from the
upstream region to the reconnection exhaust during plasma convection.
Note that if the frozen-in flux condition between the plasma and magnetic field is
satisfied, the condition of $\vec{E} + \vec{v} \times \vec{B}/c=0$
is equivalent to the equation
$(\partial /\partial t + \vec{v} \cdot \nabla) A_z =0$
 in two-dimensional system. 
Then the plasma quantities associated with the plasma flow motion as 
Lagrangian flow motion can be determined from the contour of the vector
potential $A_z$.

Two flux tubes at $t \Omega_{ci}=24.3$ are located in the upstream region and are not yet
reconnected.  The ion and electron temperatures remain low.
At $t \Omega_{ci}=31.1$, the flux tubes are just reconnecting, and they surround
both the X-type diffusion region and the boundary region of the magnetic
separatrix.  Both ions and electrons are heated in the diffusion region,
and the hot plasma is ejected outward.  At $t \Omega_{ci}=37.9$, outside the margin
of the two flux tubes, the closed flux tube starts to shrink 
and surrounds the O-type point.  The plasma in the shrinking flux
tube is compressed, and its temperature increases further.  This time
evolution is the same as that observed in other PIC simulations performed by many
researchers to date \citep[e.g.,][]{Hoshino98}.

\begin{figure}
\plotone{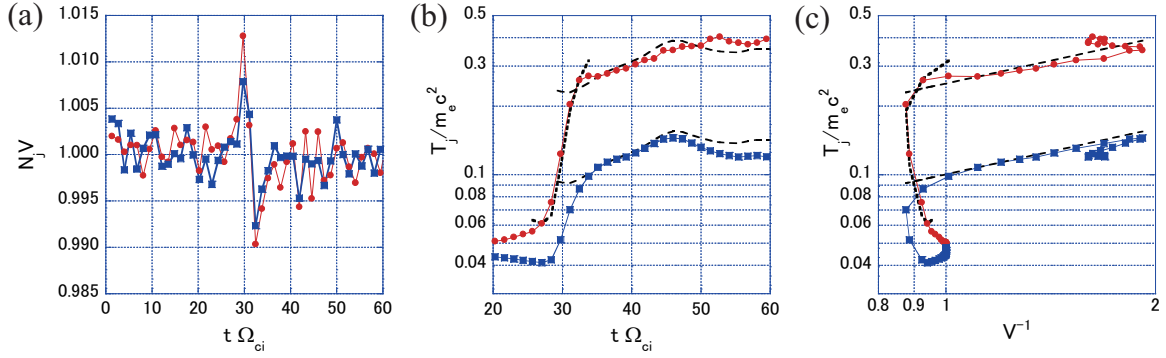}
\caption{(a) Time history of the total particle numbers of ions (red circles)
and electrons (blue squares) integrated over the volume of the magnetic flux tube
shown in Figure 1.  (b) Time history of ion (red) and electron (blue)
temperatures.  The temperatures are normalized by the electron rest mass energy.
The dashed lines after $t\Omega_{ci}=32$ are the adiabatic relation with
$T_j V^{2/3}=const.$, and the dotted line between $27 < t \Omega_{ci} < 32$
is obtained using the effective Ohmic  heating model.
(c) Phase diagram for the
reciprocal of the flux volume and the temperatures for ions (red) and electrons
(blue).  The dashed and dotted lines have the same meaning as those in (b).}
\end{figure}

Figure 2a shows the time evolution of the total particle number of ions (red circles)
and electrons (blue squares) in the flux tube in Figure 1.  The number is
normalized by the average number during reconnection.  We can see that the particle
numbers for ions and electrons are almost conserved within an error of approximately $1 \%$, and
we find that the frozen-in condition between the plasma and magnetic field lines is almost
satisfied.  However, as we expected, a slightly large departure appears in
the stage where $t \Omega_{ci} =27-33$, which corresponds to slippage
of the magnetic field lines during the passage of the flux tube through the diffusion
region.  We can also observe that the slippage of ions, which have larger gyroradii, is
slightly larger than that of electrons. The frozen-in condition of the plasma and
magnetic field lines is not perfectly satisfied, but roughly speaking, both
ions and electrons are moving together with the motion of the flux tube,
probably because the effective magnetic Reynolds number is large in collisionless
reconnection.  From this result, we can use the motion of the magnetic flux
tube as a tracer of the Lagrangian plasma flow.  

Figure 2b shows the time history of the ion and electron temperatures averaged over
the flux tube shown in Figure 1.  The temperature is calculated by taking
the second-order moment of the velocity distribution function in the simulation
frame (i.e., in the center of the flux tube frame), 
and our temperature includes the bulk flow energy, because we study
the gross kinetic energy contained inside flux tube. 
Figure 2c is a phase diagram of the relationship between their temperatures
and the volume of the flux tube $V$, which is defined as the area sandwiched
between the magnetic field lines.  Because the ion and electron particle numbers
in the flux tube are unchanged from those in Figure 2a, taking the reciprocal of
the volume, $V^{-1}$ yields the averaged plasma density for the flux tube.
We normalized the volume of the flux tube by that in the initial state.

In the initial state of $t \Omega_{ci} = 0$ in Figure 2c, both ions and
electrons are situated at $(T,V^{-1})=(0.05,1)$, and the flux tube is located
outside the plasma sheet.  With increasing time, but still in the early stage 
($t\Omega_{ci}<27$) in Figures 2b and 2c, both the ions and electrons remain cold,
but the electrons are slightly cooler, probably because of adiabatic
expansion of the magnetized electrons.  As the flux tube is convected
toward the plasma sheet, the outward-propagating
slow expansion wave from the X-point causes the flux tube to widen
(see the widening of the flux tube above and below the X-type point at
$t \Omega_{ci}=24.3$ in Figure 1).  

When the flux tube is crossing the diffusion region at $t \Omega_{ci}=27-33$,
which is the time at which the flux tube exhibits slippage in Figure 2a,
both the ion and electron temperatures start to increase rapidly.  The
electron temperature doubles, whereas the ion temperature becomes almost four
times larger.  From Figure 2c, we find that the volume of the flux tube does
not change very much.  The increase in plasma temperature is independent of
the flux tube compression, suggesting that the energy gain originated in
a nonadiabatic process, probably effective Ohmic  heating by
collisionless inertial resistivity.

For $t \Omega_{ci} > 35$, i.e., after the flux tube
has finished crossing the diffusion region, both ions and electrons are gradually heated
as the flux tube is transported outward.  We found that 
the gradual increases in the temperatures $T_i$ and $T_e$ are almost proportional
to $1/V^{\gamma-1}$ with $\gamma=5/3$, as denoted by the black dashed lines
in Figures 2b and 2c.  This suggests that the adiabatic heating process dominates
both ions and electrons after the two flux tubes merge.
Although the plasma temperature inside the flux tube is not uniform in space, it is
interesting to note that the temperature averaged over the flux tube satisfies the
relation describing the adiabatic process.

At a later time stage, we can see that the temperature oscillates.
Owing to the periodic boundary in $x$, the high-pressure plasmoid in and around
the O-type point pushes the reconnection outflow backward, and compression and
expansion of the high-pressure plasmoid can be seen.
From Figures 2b and 2c, we may conclude that the reconnection heating can be
clearly separated into two stages, the nonadiabatic process during the magnetic field reconnection stage and the adiabatic process after 
reconnection.  
Note that during the adiabatic process, the ratio of the ion temperature to the electron
temperature does not change, and thus the final energy partition seems to be
determined solely during the nonadiabatic process. 

\begin{figure}
\plotone{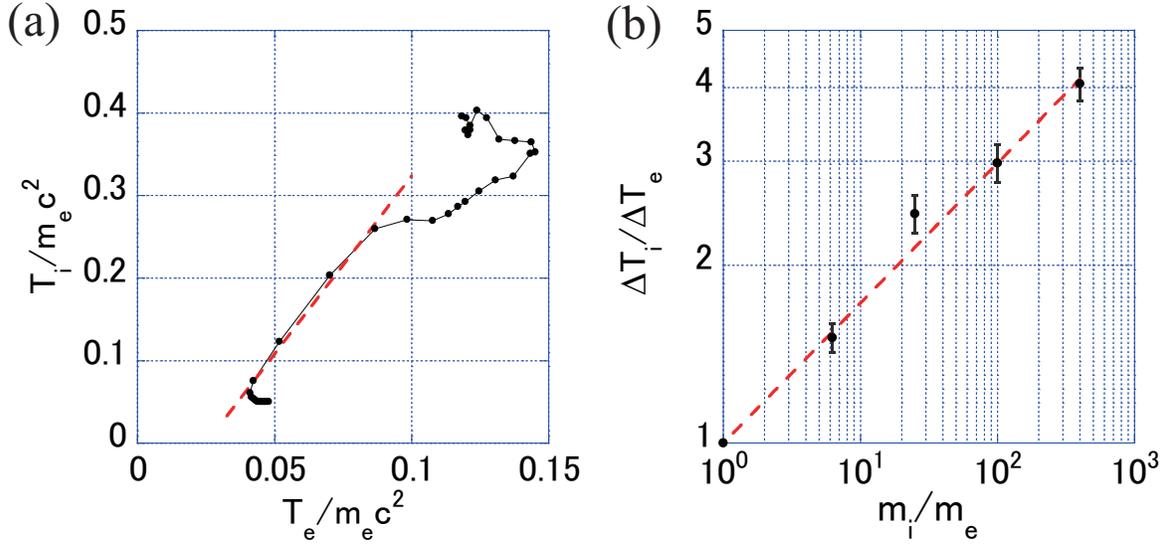}
\caption{(a) Time history of ion and electron temperatures integrated along the
magnetic flux tube.  The red line is the best fit for the nonadiabatic
heating stage.  (b) Mass dependence of the ratio of the increment
 of the ion temperature to the electron temperature.
Red line is the power law fitting $(m_i/m_e)^p$ with $p=0.237$.}
\end{figure}

Let us discuss the preferential heating during the nonadiabatic stage.
Figure 3a shows the relationship between the ion and electron temperatures during
magnetic reconnection; we focus on the nonadiabatic process stage for
$t \Omega_{ci}=27.0-32.4$.  The increase in the ion temperature is
almost proportional to the increase in the electron one, and the red line is
a linear fitting line with a slope of $4.31$.  In addition to analyzing the typical
flux tube shown in Figure 1, we performed the same analysis for other magnetic
flux tubes in Figure 1 and found that the ratio of the increment of the ion temperature to that of the electron
temperature, $\Delta T_i/\Delta T_e$, is approximated by $4.06 \pm 0.26$.   

The energy partition between ions and electrons should depend on the mass ratio of
ions to electrons artificially assumed in the PIC simulation. Therefore, in addition to the PIC
simulation of the mass ratio $m_i/m_e=400$ in Figure 1, we performed other simulation
runs with $m_i/m_e= 6.25$, $25$ and $100$.  Figure 3b shows the
dependence of the energy partition on the mass ratio.  If we can fit the mass
dependence by a power law function of $m_i/m_e$, the best fit can be approximated by
\begin{equation}
  \frac{\Delta T_i}{\Delta T_e}
  = \left( \frac{m_i}{m_e} \right)^{0.237 \pm 0.005}.
\end{equation}
Note that we assumed $\Delta T_i/\Delta T_e =1$ for the pair plasma.

According to this scaling law, we will be able to estimate the increment of
the ion temperature for a given electron temperature. The dotted lines in
the nonadiabatic stage in Figures 2b and 2c are obtained by assuming
$\Delta T_i/\Delta T_e = (m_i/m_e)^{1/4}=4.47$.
We find that the dotted lines are almost superposed on the ion temperature curves.

\section{Nonadiabatic Heating Process} 
Let us discuss theoretically the nonadiabatic heating process. 
We think that there are two important heating processes involved in this stage:
one is energy dissipation in the magnetic diffusion region, which is
provided by the effective Ohmic heating under the collisionless inertial
resistivity. The other is kinetic heating in the magnetic separatrix,
namely, the boundary between the upstream region of the plasma sheet and
the plasma outflow region with reconnection jet flows. 

We first estimate the energy gain in the diffusion region.
The effective Ohmic heating rates $Q_j$ of ions and electrons
integrated over the ion/electron diffusion region
can be estimated
as $Q_j = E J_j S_j$, where $E$, $J_j$, and $S_j$ are the reconnection electric
field, ion/electron electric current, and volume of the ion/electron
diffusion region, respectively.  The volume of the diffusion region in a two-dimensional
system, $S_j=\Delta_j d_j$, may be simply assumed to be a rectangular box with
$\Delta_j$ in the $x$ direction and $d_j$ in the $y$ direction.

In collisionless reconnection, $d_j$ can be easily evaluated using the
bounce motion across the neutral sheet, which is the so-called meandering
particle orbit, where the local gyroradius becomes comparable to the distance
$d_j$ from the X-type neutral point, and $\Delta_j$ is the gyroradius of
the reconnected magnetic field described by the Speiser orbit.  Namely,
$d_j=\sqrt{\lambda_y v_{j,in}/\Omega_{j,in}}$,
and
$\Delta_j =\sqrt{\lambda_x v_{j,out}/\Omega_{j,out}}$,
where $\lambda_x$ and $\lambda_y$ are the scale lengths of the magnetic field
gradient in the $x$ and $y$ directions, respectively. Further,
$v_{j,out}/\Omega_{j,out}$ and $v_{j,in}/\Omega_{j,in}$ are the gyroradius
defined by the outflow/inflow velocity and the gyrofrequency based on the
asymptotic magnetic field strength in the outflow/inflow region, respectively
 \citep[e.g.,][]{Coroniti85,Buechner89}. 

The ion/electron electric current inside the diffusion region can be evaluated 
using Ohm's law, $J_j=\sigma_j E$, where $\sigma_j=ne^2/m_j \nu_j$ is the effective
electric conductivity.  For collisionless reconnection, the effective
collision frequency, $\nu_j = v_{j,out}/\Delta_j$, can be estimated from the finite
residence time in the magnetic diffusion region, because the resonance between
the reconnection electric field $E$ and the particle acceleration due to $E$
is the origin of the inertial resistivity in collisionless reconnection
\citep[e.g.,][]{Coppi66,Hoh66}.

In the inflow region, the plasmas are convected by $E \times B$ drift motion,
whose speed is much smaller than the Alfv\'en speed, defined by
the magnetic field in the asymptotic field strength upstream and the plasma
density in the plasma sheet. As we assumed that the background plasma temperature
upstream is the same as that in the plasma sheet, the Alfv\'en speed is
almost the same as the ion thermal velocity.   As long as the ion and
electron thermal velocities are larger than $E \times B$ drift velocity,
$v_{j,in}$ for the meandering motion  
may be estimated to be their thermal velocities of
$\sqrt{2 T_{j0}/m_j}$, respectively.

Using the above relations, we can obtain,
\begin{displaymath}
  \frac{J_i \Delta_i}{J_e \Delta_e}=\frac{\sigma_i \Delta_i}{\sigma_e \Delta_e}
  =\left( \frac{m_e v_{e,out}}{m_i v_{i,out}} \right)
    \left(\frac{\Delta_i}{\Delta_e} \right)^2 = 1,
\end{displaymath}
and
\begin{displaymath}
  \frac{d_i}{d_e}=\left( \frac{v_{i,in} m_i}{v_{e,in} m_e} \right)^{1/2} 
            = \left(\frac{m_i T_{i0}}{m_e T_{e0}}\right)^{1/4}.
\end{displaymath}
Then the ratio of the effective Ohmic heating of ion to electron integrated over
the magnetic diffusion region and the raito of the increment of temperatures
averaged over the flux tube are given by
\begin{equation}
  \frac{Q_i}{Q_e} = \frac{\Delta T_i}{\Delta T_e}
      = \frac{J_i E S_i}{J_e E S_e} = \frac{\sigma_i \Delta_i d_i}{\sigma_e \Delta_e d_e}
     = \left(\frac{m_i T_{i0}}{m_e T_{e0}}\right)^{1/4}.
\end{equation}
This result is in good agreement with the numerical simulation result in
Equation (1) during the non-adiabatic heating stage. 

As one can see the above theoretical argument,
our estimation does not explicitly include the nonadiabatic
heating process at the separatrix boundary, but we implicitly took
account of the energy gain at the boundary due to the
plasma beam population/heat flux emanating from the diffusion region.
As long as the distance is not far from the diffusion region,
the plasma heating at the separatrix would be provided through the beam
plasma instability between the cold component transported by $E \times B$
drift motion from the upstream and the beam component coming from the
diffusion region \citep[e.g.,][]{Hoshino98}, and the free energy of the beam
plasma instability is mainly carried by the beam component 
\citep[e.g.,][]{Lottermoser98,Hoshino98}.

At the separatrix boundary far from the diffusion region, the contribution
from the local plasma heating by the pick-up plasma process across the boundary
may dominate, but our simulation size is just several times larger as compared
to the size of the magnetic diffusion region.  In fact, the plasma
sheet is known to be unstable to the tearing mode instability
\citep[e.g.,][]{Coppi66,Hoh66}, 
and the separation distance between two X-type points may not be large
compared to the size of diffusion region. \citet{Sironi14} shows the
sporadic formation of many plasmoids in the plasma sheet.
For such a plasmoid-dominated reconnection, 
our theoretical argument based on the magnetic-diffusion-dominated,
nonadiabatic heating seems to be valid.

\section{Summary and Discussion} 

We discussed the time evolution of ion and electron temperature integrated
along the time evolving magnetic flux tube for a plasma sheet with the
anti-parallel magnetic field topology, and found that the plasma heating
process during reconnection can be separated into two stages of adiabatic
and non-adiabatic processes.
Except for the time stage of two flux tube merging, the heating
process can be approximated by the adiabatic heating from the macroscopic
point of view.  We also found in the nonadiabatic heating stage of two flux
tube merging the increment of ion and electron temperature ratio is
proportional to $(m_i/m_e )^{1/4}$.

Our simulation study in two-dimensional system with the anti-parallel
magnetic field topology is only a first step toward understanding the energy
partition between ion and electron in a variety of astroplasma environments.
It is important to further examine our simple model of adiabatic vs
nonadiabatic heating in more general cases.  
For a case of non-anti-parallel magnetic field topology with a finite guide
field, the mechanism of the collisionless energy dissipation in the magnetic
diffusion region would be different from our theoretical argument in Equation (2)
\citep{Drake77}.  In three-dimensional system, the lower-hybrid drift
instability would be also expected to contribute to 
electron heating in the plasma sheet boundary in three-dimensional system
\citep{Davidson75,Shinohara01}.  
Recent three-dimensional PIC simulations with a finite guide
field show the diffusion region becomes highly dynamical state in association
with a spontaneous formation of the flux rope \citep{Daughton11,Liu13}. 
A study of the plasma heating in a three-dimensional and larger scale system
would be especially important.

In the paper we discussed the energy partition between two compositions of ion
and electron, but our findings may have important implications to the
energy partition of multi-component plasmas including heavy ions observed
in solar flares and in the earth's substorms. \\

This work was supported by JSPS Grant-in-Aid for Scientific Research
(KAKENHI) Grant No. 18K18748.  The author thanks T. Amano,
W. Baumjohann, J. B\"uchner, C. C. Haggerty, M. Hesse, H. Ji,
R. Nakamura, Y. Ohira, A. Vivads and L. M. Zelenyi
for valuable discussions.


\begin{thebibliography}{}
\expandafter\ifx\csname natexlab\endcsname\relax\def\natexlab#1{#1}\fi
\providecommand{\url}[1]{\href{#1}{#1}}
\providecommand{\dodoi}[1]{doi:~\href{http://doi.org/#1}{\nolinkurl{#1}}}
\providecommand{\doeprint}[1]{\href{http://ascl.net/#1}{\nolinkurl{http://ascl.net/#1}}}
\providecommand{\doarXiv}[1]{\href{https://arxiv.org/abs/#1}{\nolinkurl{https://arxiv.org/abs/#1}}}

\bibitem[{{Baumjohann} {et~al.}(1989){Baumjohann}, {Paschmann}, \&
  {Cattell}}]{Baumjohann89}
{Baumjohann}, W., {Paschmann}, G., \& {Cattell}, C.~A. 1989, \jgr, 94, 6597,
  \dodoi{10.1029/JA094iA06p06597}

\bibitem[{{Birn} {et~al.}(2012){Birn}, {Artemyev}, {Baker}, {Echim}, {Hoshino},
  \& {Zelenyi}}]{Birn12}
{Birn}, J., {Artemyev}, A.~V., {Baker}, D.~N., {et~al.} 2012, \ssr, 173, 49,
  \dodoi{10.1007/s11214-012-9874-4}

\bibitem[{{Blandford} {et~al.}(2017){Blandford}, {Yuan}, {Hoshino}, \&
  {Sironi}}]{Blandford17}
{Blandford}, R., {Yuan}, Y., {Hoshino}, M., \& {Sironi}, L. 2017, \ssr, 207,
  291, \dodoi{10.1007/s11214-017-0376-2}

\bibitem[{{B{\"u}chner} \& {Zelenyi}(1989)}]{Buechner89}
{B{\"u}chner}, J., \& {Zelenyi}, L.~M. 1989, \jgr, 94, 11821,
  \dodoi{10.1029/JA094iA09p11821}

\bibitem[{{Cerutti} {et~al.}(2013){Cerutti}, {Werner}, {Uzdensky}, \&
  {Begelman}}]{Cerutti13}
{Cerutti}, B., {Werner}, G.~R., {Uzdensky}, D.~A., \& {Begelman}, M.~C. 2013,
  \apj, 770, 147, \dodoi{10.1088/0004-637X/770/2/147}

\bibitem[{{Coppi} {et~al.}(1966){Coppi}, {Laval}, \& {Pellat}}]{Coppi66}
{Coppi}, B., {Laval}, G., \& {Pellat}, R. 1966, Physical Review Letters, 16,
  1207, \dodoi{10.1103/PhysRevLett.16.1207}

\bibitem[{{Coroniti}(1985)}]{Coroniti85}
{Coroniti}, F.~V. 1985, \jgr, 90, 7427, \dodoi{10.1029/JA090iA08p07427}

\bibitem[{{Daughton} {et~al.}(2011){Daughton}, {Roytershteyn}, {Karimabadi},
  {Yin}, {Albright}, {Bergen}, \& {Bowers}}]{Daughton11}
{Daughton}, W., {Roytershteyn}, V., {Karimabadi}, H., {et~al.} 2011, Nature
  Physics, 7, 539, \dodoi{10.1038/nphys1965}

\bibitem[{{Davidson} \& {Gladd}(1975)}]{Davidson75}
{Davidson}, R.~C., \& {Gladd}, N.~T. 1975, Physics of Fluids, 18, 1327,
  \dodoi{10.1063/1.861021}

\bibitem[{{Decoster} \& {Frank}(1979)}]{DeCoster79}
{Decoster}, R.~J., \& {Frank}, L.~A. 1979, \jgr, 84, 5099,
  \dodoi{10.1029/JA084iA09p05099}

\bibitem[{{Drake} \& {Lee}(1977)}]{Drake77}
{Drake}, J.~F., \& {Lee}, Y.~C. 1977, Physics of Fluids, 20, 1341,
  \dodoi{10.1063/1.862017}

\bibitem[{{Drake} {et~al.}(2009){Drake}, {Swisdak}, {Phan}, {Cassak}, {Shay},
  {Lepri}, {Lin}, {Quataert}, \& {Zurbuchen}}]{Drake09}
{Drake}, J.~F., {Swisdak}, M., {Phan}, T.~D., {et~al.} 2009, Journal of
  Geophysical Research (Space Physics), 114, A05111,
  \dodoi{10.1029/2008JA013701}

\bibitem[{{Eastwood} {et~al.}(2013){Eastwood}, {Phan}, {Drake}, {Shay}, {Borg},
  {Lavraud}, \& {Taylor}}]{Eastwood13}
{Eastwood}, J.~P., {Phan}, T.~D., {Drake}, J.~F., {et~al.} 2013, Physical
  Review Letters, 110, 225001, \dodoi{10.1103/PhysRevLett.110.225001}

\bibitem[{{Egedal} {et~al.}(2013){Egedal}, {Le}, \& {Daughton}}]{Egedal13}
{Egedal}, J., {Le}, A., \& {Daughton}, W. 2013, Physics of Plasmas, 20, 061201,
  \dodoi{10.1063/1.4811092}

\bibitem[{{Haggerty} {et~al.}(2015){Haggerty}, {Shay}, {Drake}, {Phan}, \&
  {McHugh}}]{Haggerty15}
{Haggerty}, C.~C., {Shay}, M.~A., {Drake}, J.~F., {Phan}, T.~D., \& {McHugh},
  C.~T. 2015, \grl, 42, 9657, \dodoi{10.1002/2015GL065961}

\bibitem[{{Hesse} {et~al.}(2001){Hesse}, {Birn}, \& {Kuznetsova}}]{Hesse01}
{Hesse}, M., {Birn}, J., \& {Kuznetsova}, M. 2001, \jgr, 106, 3721,
  \dodoi{10.1029/1999JA001002}

\bibitem[{{Hoh}(1966)}]{Hoh66}
{Hoh}, F.~C. 1966, Physics of Fluids, 9, 277, \dodoi{10.1063/1.1761670}

\bibitem[{{Hoshino} \& {Lyubarsky}(2012)}]{Hoshino12}
{Hoshino}, M., \& {Lyubarsky}, Y. 2012, \ssr, 173, 521,
  \dodoi{10.1007/s11214-012-9931-z}

\bibitem[{{Hoshino} {et~al.}(2001){Hoshino}, {Mukai}, {Terasawa}, \&
  {Shinohara}}]{Hoshino01}
{Hoshino}, M., {Mukai}, T., {Terasawa}, T., \& {Shinohara}, I. 2001, \jgr, 106,
  25979, \dodoi{10.1029/2001JA900052}

\bibitem[{{Hoshino} {et~al.}(1998){Hoshino}, {Mukai}, {Yamamoto}, \&
  {Kokubun}}]{Hoshino98}
{Hoshino}, M., {Mukai}, T., {Yamamoto}, T., \& {Kokubun}, S. 1998, \jgr, 103,
  4509, \dodoi{10.1029/97JA01785}

\bibitem[{{Imada} {et~al.}(2005){Imada}, {Hoshino}, \& {Mukai}}]{Imada05}
{Imada}, S., {Hoshino}, M., \& {Mukai}, T. 2005, \grl, 32, L09101,
  \dodoi{10.1029/2005GL022594}

\bibitem[{{Liu} {et~al.}(2013){Liu}, {Daughton}, {Karimabadi}, {Li}, \&
  {Roytershteyn}}]{Liu13}
{Liu}, Y.-H., {Daughton}, W., {Karimabadi}, H., {Li}, H., \& {Roytershteyn}, V.
  2013, Physical Review Letters, 110, 265004,
  \dodoi{10.1103/PhysRevLett.110.265004}

\bibitem[{{Lottermoser} {et~al.}(1998){Lottermoser}, {Scholer}, \&
  {Matthews}}]{Lottermoser98}
{Lottermoser}, R.-F., {Scholer}, M., \& {Matthews}, A.~P. 1998, \jgr, 103,
  4547, \dodoi{10.1029/97JA01872}

\bibitem[{{Lyubarsky} \& {Kirk}(2001)}]{Lyubarsky01}
{Lyubarsky}, Y., \& {Kirk}, J.~G. 2001, \apj, 547, 437, \dodoi{10.1086/318354}

\bibitem[{{Phan} {et~al.}(2013){Phan}, {Shay}, {Gosling}, {Fujimoto}, {Drake},
  {Paschmann}, {Oieroset}, {Eastwood}, \& {Angelopoulos}}]{Phan13}
{Phan}, T.~D., {Shay}, M.~A., {Gosling}, J.~T., {et~al.} 2013, \grl, 40, 4475,
  \dodoi{10.1002/grl.50917}

\bibitem[{{Shay} {et~al.}(2018){Shay}, {Haggerty}, {Matthaeus}, {Parashar},
  {Wan}, \& {Wu}}]{Shay18}
{Shay}, M.~A., {Haggerty}, C.~C., {Matthaeus}, W.~H., {et~al.} 2018, Physics of
  Plasmas, 25, 012304, \dodoi{10.1063/1.4993423}

\bibitem[{{Shay} {et~al.}(2014){Shay}, {Haggerty}, {Phan}, {Drake}, {Cassak},
  {Wu}, {Oieroset}, {Swisdak}, \& {Malakit}}]{Shay14}
{Shay}, M.~A., {Haggerty}, C.~C., {Phan}, T.~D., {et~al.} 2014, Physics of
  Plasmas, 21, 122902, \dodoi{10.1063/1.4904203}

\bibitem[{{Shinohara} {et~al.}(2001){Shinohara}, {Suzuki}, {Fujimoto}, \&
  {Hoshino}}]{Shinohara01}
{Shinohara}, I., {Suzuki}, H., {Fujimoto}, M., \& {Hoshino}, M. 2001, Physical
  Review Letters, 87, 095001, \dodoi{10.1103/PhysRevLett.87.095001}

\bibitem[{{Sironi} \& {Spitkovsky}(2014)}]{Sironi14}
{Sironi}, L., \& {Spitkovsky}, A. 2014, \apjl, 783, L21,
  \dodoi{10.1088/2041-8205/783/1/L21}

\bibitem[{{Spruit}(2010)}]{Spruit10}
{Spruit}, H.~C. 2010, in Lecture Notes in Physics, Berlin Springer Verlag, Vol.
  794, Lecture Notes in Physics, Berlin Springer Verlag, ed. T.~{Belloni}, 233

\bibitem[{{Uzdensky}(2011)}]{Uzdensky11}
{Uzdensky}, D.~A. 2011, \ssr, 160, 45, \dodoi{10.1007/s11214-011-9744-5}

\bibitem[{{Wang} {et~al.}(2012){Wang}, {Gkioulidou}, {Lyons}, \&
  {Angelopoulos}}]{Wang12}
{Wang}, C.-P., {Gkioulidou}, M., {Lyons}, L.~R., \& {Angelopoulos}, V. 2012,
  Journal of Geophysical Research (Space Physics), 117, A08215,
  \dodoi{10.1029/2012JA017658}

\end{thebibliography}


\end{document}